\documentclass[twocolumn]{article} 

\usepackage{graphicx}
\usepackage{hyperref}
\usepackage{twoopt}
\usepackage{authblk}
\newcommand{\ackname}{Acknowledgements}

\begin{document}
\title{United by skies, divided by language -- astronomy 
publishing in languages with small reader base}

\author[1]{Valentin D. Ivanov}
\affil[1]{\small European Southern Observatory, Karl-Schwarzschild-Str. 2,
85748 Garching bei M\"unchen, Germany; vivanov@eso.org}
\date{2022-06-29}

\maketitle

\begin{abstract}
\begin{small}
% \vspace{-3mm}
The mysteries of the Universe are international, the skies are 
not crossed by borders. However, the knowledge is transmitted by 
language, imposing linguistic barriers that are often difficult 
to break through. Bulgaria is considered as an example of a 
country with relatively small reader base -- it has a population 
of about 6.5 million (2021) and the Bulgarian language has 
probably $\sim$7 million speakers, if the diaspora in US, Germany 
and elsewhere is accounted for. The smaller-scale market, in 
comparison with larger non-English speaking countries, poses a 
number of limitation to the publishing landscape: 
(i) the local authors are discouraged to pen both popular and 
scientific astronomy books, because of the limited financial 
incentive; 
(ii) the market is heavily dominated by translations (from 
Russian before 1989, from English nowdays), but even those are 
fewer than in bigger countries, because the translation overhead 
costs are spread over smaller print runs. 
The history of the astronomy publishing in Bulgaria is summarized, 
with some distinct periods: pre-1944, the communist era 1944-1989, 
the modern times post 1989. A few notable publications are 
reviewed. 
Finally, some practices to help astronomy book publishing in 
languages with smaller reader bases are suggested, taking 
advantage of the recent technological development and of the 
relatively new paradigm of open access publications.
\vspace{-5mm}
\end{small}
\end{abstract}
\vspace{1cm}

\section{Introduction}

Diversity is a multifaceted concept. The obvious, most often 
considered dividing lines or bridges that connect people -- 
depending on the point of view -- are those of race, gender and
disability. There are good reasons for selecting this subset, 
these are important issues with significant impact on the lives
of a large number of people. Enormous efforts are put into 
understanding and crossing these dividing lines.

People may look the same, until they open their mouths and speak 
up, and then some notable, to put it mildly, differences become 
apparent. Language is yet another dividing line or a bridge 
that is worth considering in this aspect, because it is a primary 
venue of communication in the modern connected and globalized 
world. Modern astronomy has trespassed the national borders, the 
large collaborations that move the forefront of our science 
today are largely international and they need efficient 
communication.

The linguistic environment has two aspects. First, the 
professional scientists are ``required'' to speak English -- 
the Latin of today -- at reasonably good level. The pressures 
of publishing, participation in international conferences and 
access to the literature make this unavoidable. There can be 
only one best practice -- invest time and effort to learn and 
improve your English reading and writing skills.

However, the language also has impacts on the general acceptance 
of science in the society. The high school education is one of 
the leading factors for building a science friendly environment 
\cite{2018arXiv181201582I}. Another factor is the outreach 
activity of the astronomical departments and observatories 
in the social media and the friendly collaboration with the more 
traditional news outlets.

It is becoming increasingly more difficult today to separate 
the national and international social and news media. True, 
some are split into English, French, Chinese, Russian, etc. 
``zones'' but the young generations grow increasingly 
``latinized'' and gather most of their information from the 
English segments. This is somewhat subjective statement, based 
on a personal observation of the author: the links posted 
today in various Bulgarian forums overwhelmingly point towards 
content in English, most notably the English Wikpedia. 
This observation can easily be rationalized: the English 
sources contain more variety (the staggering differences in 
the number of Wikipedia entries in English and in Bulgarian is 
an evidence for that $\sim$6.6 vs. $\sim$0.3 million,
respectively, as of late-2022) and they are faster in 
delivering the news. 
There are probably many reasons for this state of the affairs. 
The most critical of them is probably the advantage of scale -- 
the wider (and generally richer, because it is dominated by 
first world countries) readers' base ensures that they have 
higher operational budgets than the alternatives, and they 
attract the best talent (because they have a big talent pool 
to choose from and because they can afford more competitive 
remuneration to attract that best talent, too).

\section{A case study: astronomy in Bulgaria}

Here Bulgaria is considered as an example of a country with 
relatively limiter reader base, when it comes to popular 
astronomy books (and to any books, in fact): as of 2021 it 
has a population of about 6.5 
million\footnote{\url{https://www.nsi.bg/en/content/19807/nsi-announces-final-results-population-number-bulgaria}} 
and the Bulgarian language has probably $\sim$7 million 
speakers, if the diaspora in US, Germany and elsewhere is 
accounted for. Other countries speak similar languages, but
the similarities are not close enough to allow a common book
market; a Westerner can think of the similarity between 
Spanish and Italian or Portuguese -- people may be able to 
understand each other in informal communication, but would 
find it challenging to follow dense scientific or educational 
texts.

More information on Bulgarian astronomy can be found in 
\cite{2011BlgAJ..15..129K} who reviews its ancient history and
in \cite{1999POBeo..64...45I,2018A&AT...30..441P} who summarize 
its more recent development.

\section{Astronomy books in Bulgaria}

The data in this analysis come from two sources, described in 
the next two subsections and reflect their state as of mid-June
2022.

\subsection{List of astronomy books of the Public Astronomical 
Observatory and Planetarium ``Jordano Bruno'' -- Dimitrovgrad}

The Public Astronomical Observatory and Planetarium ``Jordano 
Bruno'' -- Dimitrovgrad opened in 1962 and has been one of the 
leading centers of high-school educational and and outreach 
activities in the country ever 
since\footnote{More information about the Observatory is available 
on its website at \url{https://www.naopjbruno.bg/?page_id=248&lang=en}.} 
Its staff maintains an on-line list of Bulgarian astronomy 
books with 181 unique titles (183 
editions)\footnote{\url{https://www.naopjbruno.bg/?page_id=248&lang=en}}.
For comparison, the Library of the US Congress lists 10,000 
astronomy books at which point the search results are
truncated.

This list is incomplete -- is misses some recent books; this 
should not be taken as a major criticism, though; collecting 
all this information and maintaining it current is a gargantuan
task.

\begin{figure*} %[!htb]
\centering
\includegraphics[width=16.2cm]{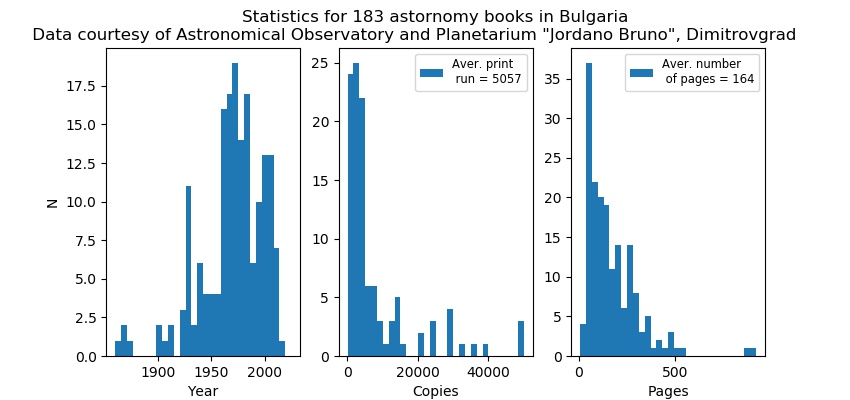}
\caption{Statistics for the books on the list from the Public 
Astronomical Observatory and Planetarium ``Jordano Bruno'': 
number of books per year (left), distribution of the number of 
printed copies (middle) and book volumes 
(right).}\label{fig:astro_book_hist}
\end{figure*} 

Figure\,\ref{fig:astro_book_hist} shows some statistics for the 
books on the list: number of books per year (left), distribution 
of the number of printed copies (middle) and book volumes (right).
The number of books has increased steadily until mid-XX century
and then there was an explosion, probably because the communist 
rulers of Bulgaria at the time perceived astronomy as a tool of
anti-religious propaganda and because the STEM disciplines (even
though they were not called that back then) were highly regarded 
at the time when the country underwent a rapid industrialization.
The drop in mid/late-1980 is associated with the economic 
difficulties during the transition to a market economy, showing
that astronomers do not live in an ivory tower, indeed. A check 
of the titles indicated that the recovery in the 1990s is 
associated with the education reform that led to the appearance 
of many new textbooks and the final decline post-2000 is just a
list incompleteness.

The next panel not particularly informative, because the print 
runs are typically available only for communist era books. 
Therefore, the average of $\sim$5000 copies is heavily dominated 
by pre-Internet books. We can speculate why modern published 
prefer to omit this kind of information. Finally the 800-page 
books are probably due to typos.

\begin{figure*} %[!htb]
\centering
\includegraphics[width=16.2cm]{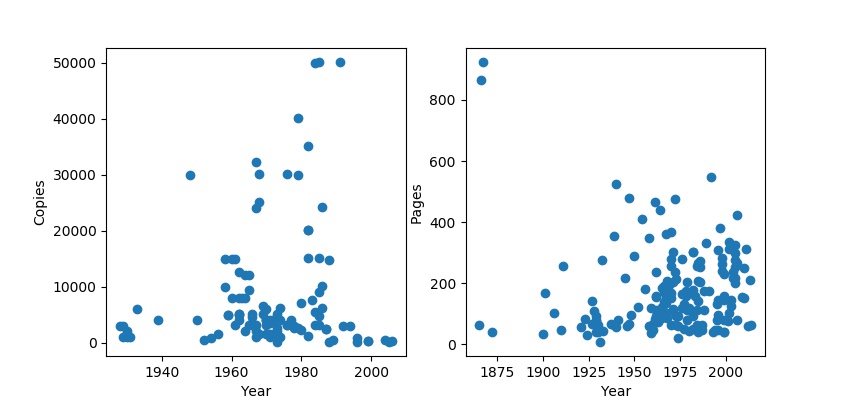}
\caption{Correlation between the the number of printed copies 
(left) and book volumes (right) and the year of publication 
for the books on the list from the Public Astronomical 
Observatory and Planetarium ``Jordano 
Bruno''.}\label{fig:astro_book_evol}
\end{figure*} 

Let us now consider the correlations between these parameters
(Fig.\,\ref{fig:astro_book_evol}). The sharp raise in the late 
1950s and early-1960s is probably associated with the advent of
the space age. Back then a considerable effort was spent 
observing artificial satellites 
\cite{1957SvA.....1..309M,1958JRASC..52...57H}, presumably for 
defense reasons. The ``secret'' printing era starting from 
1990s is now more apparent (on the left panel). The three 
books with the highest print runs are textbooks from the last
years of the communist era when the educational system was
uniformized and one-book-fitted-all. Furthermore, one of the 
three books was a Russian translation of a Bulgarian book for 
the vast USSR market. 

The structure on the book volume panel (right) reflects the 
large number of cheap and small brochures that were published 
in the years before the World War II; they were replaced by 
full-sized books in the 1950s and later.

These plots hint at the existence of three periods in the 
Bulgarian astronomy books publishing: pre-1944 when the 
country had a agriculture dominated economy, the communist 
era 1944-1989 with the rapid industrialization and the 
economic collapse at the end, and finally the modern times 
post-1989 when the country entered a semi-permanent 
political, demographic and economic turmoil. The reasons for 
that are beyond the scope of this analysis, we will only 
point that examples of Poland, Czech Republic and Slovakia, 
among others, show that this is not a necessary status or 
stage.

\subsection{List of astronomy books at the Regional Library 
``Peio Yavorov'', Burgas}

The library inventories can give a snapshot of the astronomical
literature that is available today. Furthermore, the public 
libraries in Bulgaria are nearly free, requiring only a token
membership payment, which hopefully removes the economic 
obstacles to reading and levels off the economic inequality.

The public library ``Peio Yavorov'' in 
Bourgas\footnote{\url{https://burgaslib.bg/}} was selected by 
virtue of being in the home town of the author, but there are 
also some strong reasons in terms of representation that 
motivated this choice: Bourgas with a population of just over
200,000\footnote{\url{https://www.grao.bg/tna/t41nm-15-09-2022_2.txt}} 
is the fourth most populated city in the country. Together with
about a dozen other cities with a population of over 100,000 
it hosts nearly half of the total population and probably a 
disproportionately larger fraction of students because of the
significantly older population in smaller towns. Bourgas is 
neither in the advantageous position of a county capital, not 
in the disadvantaged position of a small village with no 
industries and high unemployment.

All books tagged with ``astronomy'' were extracted from the 
library catalog, yielding 177 entries. They were classified -- 
with some degree of subjectivity -- into three categories: high 
school textbooks (93 or $\sim$53\,\%), popular science books 
(68 or $\sim$38\,\%) and science books (16 or $\sim$9\,\%).

\begin{figure*} %[!htb]
\centering
\includegraphics[width=13.2cm]{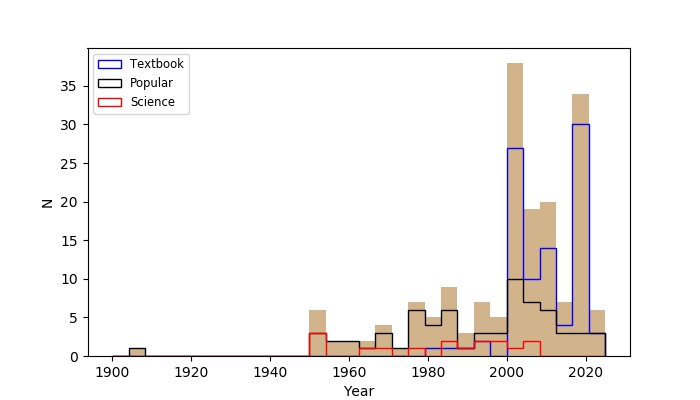}
\includegraphics[width=13.2cm]{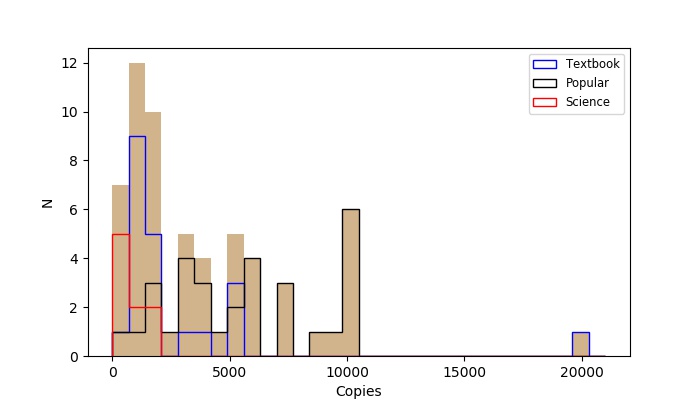}
\caption{Distributions of the astronomy books in the Regional 
Library ``Peio Yavorov'', Bourgas by year of publication (top)
and by print run (bottom): in total (beige), for high school 
textbooks (blue), for popular science books (black) and for 
science books (red).
}\label{fig:bg_astro_book_hist_0102}
\end{figure*}

The distributions of the astronomy books in the library by year 
of publication and by print run are shown in 
Fig.\,\ref{fig:bg_astro_book_hist_0102}.
The list is dominated by post-2000 books and particularly by 
high school textbooks -- easy to understand, because the books 
wear out and the libraries periodically try to update their 
inventories and to respond to the need of their most numerous 
reader base from high schools. The repeated peaks every 4-5
years correspond to the educational reforms in the school 
system that imply textbook updates and new editions. 

The oldest astronomy book in the library is a Bulgarian 
translation of Camille Flammarion ``Astronomy for Women'' and 
it is probably more important as a historic curiosity and an 
evidence of gender politics of the time than as an actual 
educational tool.

The absolute champion with the largest print run is a high 
school textbook form 1995 by N. Nikolov, a professor at the 
University of Sofia, with 20070 copies. This book appeared 
in the transition years between the communist educational 
system that required maximum uniformity and the new democratic
system that allowed selection of textbooks, which led to more 
diverse but fragmented textbook landscape. Science textbooks 
are lead by two books from I. Kozhurov (1970, 2081 copies) 
and N. Bonev (1964, 2070 copies) that appeared at the height 
of the communist industrialization when the hard science were 
exceptionally favored.

\begin{figure} %[!htb]
\centering
\includegraphics[width=3.5cm]{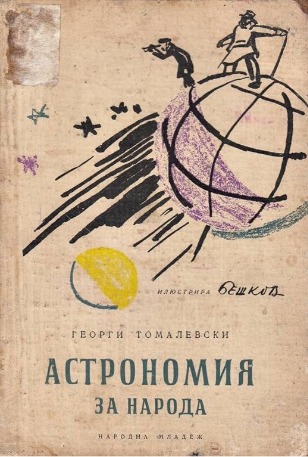}
\includegraphics[width=3.5cm]{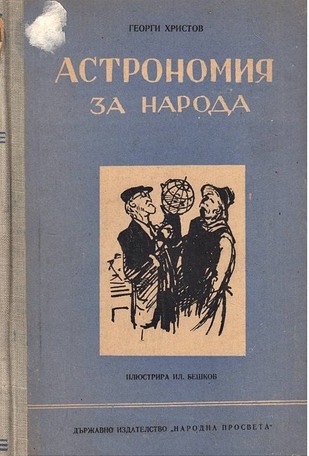}
\caption{Covers of the {\it People's astronomy} by Georgi
Hristov Tomalevski, illustrated by the famed bulgarian 
painter Ilia Beshkov.}\label{fig:tomalevki_covers}
\end{figure} 

The popular science book category is led by S. Weinberg 
(1984, 10112 copies), followed by B. Kunzetsov (1977, 10090),
B. Vorontsov-Velyaminov (1973, 10085), Y. Perelman (1961, 
10080), S. Mace (1995, 10000), G. Tomalevski (1958, 10000;
Fig.\,\ref{fig:tomalevki_covers}), 
and P. Maffei (with two entries: 1986, 9517; 1985, 8821). 
Apparently, for many decades in the Bulgarian publishing
10000 copies was considered somewhat is a ``standard'' run 
for popular astronomy books. The only ``native'' entry in 
this list is the book by Tomalevski, published 64 years ago.

The last observation prompts one to consider if the science 
writers from countries with limited reader base are facing 
a bigger difficulty writing and publishing their books than 
their counterparts from bigger countries, because the 
investment of time and effort to writhe a book in Bulgarian
or Romanian or Serbian from a writer in these countries is 
the same as for someone in the USA, UK or Australia to 
write a book in English, but the potential rewards are much 
smaller and the overheads are distributed over much smaller
number of printed copies. Of course, translation implied 
additional translation overhead costs, but also may mean 
that the world's best books become available. Finally, it 
should be noted that 
it is impossible to build a career in Bulgaria or any of 
the other countries in the list from science writing alone 
(as it about to become now in bigger countries as well, 
because of the slowing economy worldwide).

\begin{figure*} %[!htb]
\centering
\includegraphics[width=13.2cm]{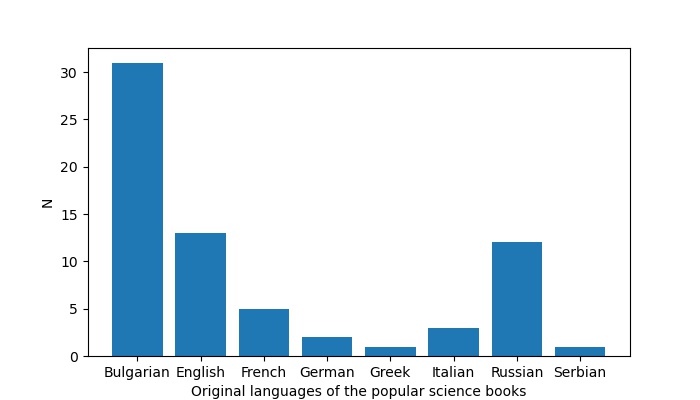}
\caption{Distributions of the astronomy books in the Regional 
Library ``Peio Yavorov'', Bourgas by language.
}\label{fig:bg_astro_book_hist_03}
\end{figure*} 

Fig.\,\ref{fig:bg_astro_book_hist_03} answers the question where
the science popular books come from -- it shows a distribution 
of their original language. Only about half of the books were
written in Bulgarian, the rest are translations. The two 
dominant languages are English and Russian, reflecting two 
tendencies: the leading role that the English language and the 
English speaking countries take in the modern globalized world, 
and the past ties of the communist Bulgaria with the USSR as 
part of the Easter Block. One can speculate that small countries 
in Africa or South America would have similar bias towards the 
languages spoken in their former colonial masters.

A further, closer look at the library inventory reveals an 
alarming result: some major popular books are missing, even 
though they are recent prints and are readily available in 
the bookstores.
Among them are titles by Michio Kaku, Stephen Houcking, Neil 
deGrasse Tyson and even by Carl Sagan. We can only speculate 
what is the reason -- if it is due to increasing book prices 
combined with reduced funds for public library or if it is a
matter of less than ideally informed book selection.

\section{Discussion and summary}

This short exploration of the astronomy book landscape in 
small countries rests upon the assumptions that (i) Bulgaria 
is a representative of this type of countries (at least as 
far as the former Eastern Europe) and that (ii) the public 
library in Bourgas -- a major bibliographical source here --
reflects the astronomy books available to the average reader
at the moment.

The first and sad conclusion is that the science books are 
expensive, even more for countries with small language bases 
(e.g., due to additional translation costs) that in addition 
usually goes hand in hand with poor economy; books are 
expensive for both libraries and individuals and there are 
notable and worrying misses in public libraries.
Next, the local authors are discouraged to pen popular and 
scientific books, because of the limited financial incentive.
The print runs are shrinking over time (admittedly, the books 
are becoming bigger). 
Finally, half of the popular book market is foreign-dominated.
This is not necessarily a criticism, because it makes 
available the worlds best astronomy books.

These observations upon the Bulgarian astronomy books can 
offer very little recommendations for improvement. Writing 
such books is very likely to remain work of a few enthusiastic
and enlightened educators and scientists; the publishes will
always have the alternative to turn to translating books from 
the  significant body of foreign works. Striving for high 
quality and making their book more accessible and adapted to 
the Bulgarian reader -- or the reader from any other country 
with a small book market -- is probably the best hope that 
``native'' science writers have. A good example is {\it 
People's astronomy} by G. Hr. Tomalevski, who attracted Ilia
Beshkov, one of the best painters and illustrators of the day, 
to work on his book. This example points at another path -- 
if the lavish illustrations were the ``visual social media''
of the day and if the success of Tomalevski is any indication, 
then the modern science writers may benefit from exploring 
the opportunities of Internet and the social media to reach a
wider audience -- hardly a new revelation.

Finally, we point at the older astronomy books as a source of
inspiration to the new science writers and evidence, that this 
complex and challenging material can successfully be conveyed
to the public.

\section*{\ackname}
This is an extended write up of a poster presented at the European 
Week of Astronomy and Space Science (EAS) held in Valencia, Spain, 
Jun 26 -- Jul 1, 2022, Special Session 34: Diversity and Inclusion 
in European Astronomy.

\end{document}